\documentclass{ws-ijmpa}
\usepackage[super,compress]{cite}
\usepackage{graphicx}
\usepackage{amssymb}
\usepackage{bm}
\usepackage{url}
\newcommand{\nn}{\nonumber}

\begin{document}
\markboth{Y.Kitadono, M.Wakamatsu, L.-P.Zou, and P.-M.Zhang}{Role of guiding center in Landau level system and mechanical and pseudo orbital angular momenta}

%
\catchline{}{}{}{}{}
%

\title{Role of guiding center in Landau level system and mechanical and pseudo orbital angular momenta}

\author{$^{1}$Yoshio Kitadono\footnote{kitadono@impcas.ac.cn (Corresponding author)}}
\address{$^{1}$Institute of Modern Physics~(IMP), Chinese Academy of Sciences~(CAS),\\
509, Nanchang Road, Lanzhou, 730000, People's Republic of China}


\author{$^{1,2}$Masashi Wakamatsu}
\address{$^2$Theory Center, Institute of Particle and Nuclear Studies (IPNS), High Energy Accelerator Research Organization (KEK),
1-1, Oho, Tsukuba, Ibaraki, 305-0801, Japan}

\author{$^{1}$Liping Zou}
\address{$^{1}$Institute of Modern Physics~(IMP), Chinese Academy of Sciences~(CAS),\\
509, Nanchang Road, Lanzhou, 730000, People's Republic of China}

\author{$^{3}$Pengming Zhang}
\address{$^3$School of Physics and Astronomy, Sun Yat-sen University,\\ Zhuhai 519082, People's Republic of China}

\maketitle

\begin{history}
\end{history}

\begin{abstract}
There is an interesting but not so popular quantity called pseudo orbital angular momentum (OAM) in the Landau-level system, besides the well-known canonical and mechanical OAMs. The pseudo OAM can be regarded as a gauge-invariant extension of the canonical OAM, which is formally gauge invariant and reduces to the canonical OAM in a certain gauge. Since both of the pseudo OAM and the mechanical OAM are gauge invariant, it is impossible to judge which of those is superior to the other solely from the gauge principle. However, these two OAMs have totally different physical meanings. The mechanical OAM shows manifest observability and clear correspondence with the classical OAM of the cyclotron motion. On the other hand, we demonstrate that the standard canonical OAM as well as the pseudo OAM in the Landau problem are the concepts which crucially depend on the choice of the origin of the coordinate system. We try to reveal the relation between the pseudo OAM and the mechanical OAM as well as their observability by paying special attention to the role of guiding-center operator in the Landau problem.
\end{abstract}

\ccode{PACS numbers: 03.65.-w; 71.70.Di; 11.90.+t}


\section{Introduction:~two gauge invariant angular momenta}
The quantum mechanics of the electron in a uniform and constant magnetic-field plays central roles to discuss the fundamental property of the electron not only in matter physics, but also in high energy physics. In particular, the electron's energy level of this system is quantized and called the Landau level \cite{Landau}. To describe this system in quantum mechanics, one needs to introduce the vector potential ${\textbf A}$ which gives the magnetic field ${\textbf B}$ through the relation, ${\textbf B}={\boldsymbol \nabla}\times {\textbf A}$.

As is known, there are two different momenta for the electron under the influence of the magnetic field.  The first is the canonical momentum, $\hat{\textbf{p}}=-i\bm{\nabla}$, and the second is the mechanical (or kinetic) momentum, $\hat{\bm{\Pi}}=\hat{\textbf{p}}+e\textbf{A}$. Obviously, the canonical momentum is not gauge-invariant operator and the mechanical momentum is gauge-invariant one. Hence the canonical one is usually believed not to describe physical observables due to its gauge dependence \cite{JJSakurai}. One can generalize this concept to other operators so that physical observables should not depend on a choice of gauge. This is called the  ``gauge principle'' and believed to work well. We can construct other operators, like orbital angular momentum~(OAM), from these canonical and mechanical momenta. Consequently, the canonical OAM, $\hat{\textbf L}^{\rm can}={\textbf x}\times \hat{\textbf p}$,  depends on a gauge and the mechanical OAM $\hat{\textbf L}^{\rm mech}={{\textbf x}\times \hat{\bm{\Pi}}}$, does not depend on the gauge. (More precise statement in quantum mechanics should be the following. The expectation values $\langle \psi|\hat{\bm{\Pi}}|\psi \rangle$ of the mechanical momentum operator $\hat{\bm{\Pi}}$ is independent of gauge choice, but the expectation value $\langle \psi|\hat{\textbf{p}}|\psi \rangle$ of the canonical momentum operator $\hat{\textbf{p}}$ is gauge dependent. With this tacit understanding, we use the simple phrase as above in the following.)

Recently, differences between the canonical OAM and the mechanical OAM for the electron in a uniform magnetic field and related topics are intensively studied in the context of a recent development of the electron vortex-beam \cite{Lcan_Nature, Bliokh_PRX, Greenshields} and angular momenta of quarks and gluon for the nucleon-spin-decomposition problem \cite{Nspin_review1, Nspin_review2}. Although the purposes to study OAMs are different, there are similarities and common interests between these two different research fields, namely, what is the difference between the canonical OAM and the mechanical OAM in quantum theory from the viewpoint of gauge theory ?

According to these recent developments on experiment and theory, it is a good time to hightlight OAMs in gauge theory again. We discussed several angular-momenta in this Landau-level system \cite{path_dep_Wakamatsu} by using DeWitt's gauge-invariant-formalism of quantum electrodynamics \cite{DeWitt} which enables us to solve the eigen equation without fixing a gauge, but with a choice of path. We focused on three OAMs; 1) the canonical OAM, 2) the mechanical OAM, and 3) ``pseudo'' OAM \cite{KM1, KM2}, where the mechanical and pseudo OAMs are gauge invariant. One can regard this pseudo-OAM as a ``gauge-invariant extension of the canonical OAM'', because it is formally gauge invariant and it reduces to the canonical OAM in a particular gauge \cite{Nspin_review1, Nspin_review2}. 

There are two questions left in our previous studies. If there is a gauge-invariant extension of a gauge-dependent operator like the pseudo OAM, the observability of this quantity is not quite clear, because the gauge principle does not necessarily forbid the observability of the pseudo OAM which is formally gauge invariant. Moreover, now we have two gauge-invariant OAMs, namely, the mechanical OAM and the pseudo OAM in the Landau-level system. What is the difference between these two gauge-invariant OAMs ? Can the pseudo OAM be an observable ?  The Landau-level system, as it is a solvable one, is a good testing-ground to answer these questions about OAMs in gauge theory.

The purpose of this paper is to clarify the difference among three OAMs in a new perspective, i.e. by paying a special attention to a reference axis for OAMs. We compare the three OAMs defined with respect to the origin and the three OAMs defined with respect to the guiding center both in classical theory and quantum theory. This comparison enables us to understand the physical meaning as well as the difference of these three OAMs in a transparent way.

The paper is organized as follows. First, in section~\ref{sec2}, we briefly review some basics of the Landau problem, wave function, energy level, and expectation values of three OAMs. Next, in section~\ref{sec3}, we address some important properties of the concept of the guiding center in classical theory and quantum theory, which is the key concept of this paper. In section~\ref{sec4}, we discuss the expectation values for the three OAMs in quantum theory to address our questions: 1) what is the difference between the mechanical and pseudo OAMs ? and 2) can the pseudo OAM be an observable because of its gauge-invariant nature ?  We summarize the paper in section~\ref{sec5}.

\section{Mechanical and pseudo orbital angular momenta of electron in a uniform magnetic field \label{sec2}}
In this section, we briefly review the well-known basics of the Landau-level system, especially by focusing on the three OAMs, i.e. the canonical, mechanical, and the pseudo OAM in this system. Throughout the paper, we discard the spin of the electron, since, within the framework of the non-relativistic treatment of the Landau problem, it can be separated from the orbital angular momentum that is of our main concern.

\subsection{Landau level system}
As is well-known, the motion of the electron with the charge $-e~(e > 0)$ and the mass $m_e$ in a uniform magnetic field is described by the Landau Hamiltonian: 
\begin{eqnarray}
 \hat{H} = \frac{\hat{\bm \Pi}^2}{2m_e} = \frac{\left(\hat{\textbf p} + e{\textbf A}\right)^2}{2m_e}, \hspace{1cm} \hat{H}\psi(x,y)=E\psi(x,y),
\end{eqnarray}
where $\hat{{\textbf p}}=-i{\bm \nabla}$ is the canonical momentum, $\hat{\bm \Pi}=\hat{\textbf p}+e{\textbf A}$ is the mechanical momentum, and ${\textbf A}$ is the gauge potential satisfying ${\textbf B} = {\bm \nabla } \times {\textbf A}$. Here, we take $z$-axis as the direction of the uniform and constant magnetic field, namely, ${\textbf B}=(0,0,B)$. We use the natural unit, $\hbar=c=1$, throughout this paper.

The eigen energy $E\equiv E_{n}$ is given by
\begin{eqnarray}
 E_{n} = \omega\left( n + \frac{1}{2} \right),  \hspace{1cm} \omega=\frac{eB}{m_e}, \label{eq.Landau.energy}
\end{eqnarray}
which is called the Landau level \cite{Landau} and $n=0,1,2,\cdots$ is the Landau quantum-number. Although the expression of the Hamiltonian depends on a choice of gauge, the Landau level does not depend on the gauge, as anticipated.
 
However, solutions of the eigen equation depend on a choice of gauge. The most convenient gauge-choice to discuss OAMs in this system is the symmetric gauge,
\begin{eqnarray}
{\textbf A}^{(\rm S)} = \frac{B}{2}\left(-y,x,0\right). \label{eq.gauge.L1.L2.S}
\end{eqnarray}
The wave function in the symmetric gauge is given in~\cite{Landau_Lifshitz}:
\begin{eqnarray}
 \psi^{(\rm S)}_{n,m}(r,\phi) 
 = N_{n,m} \frac{e^{im\phi}}{\sqrt{2\pi}} e^{-\frac{\rho}{2}} \rho^{\frac{|m|}{2}} L^{|m|}_{n-\frac{m+|m|}{2}}(\rho), \label{eq.standard.psi.S} \\
 \rho \equiv \frac{r^2}{2 l^2_B}, \hspace{1.cm}
 N_{n,m} = \frac{1}{l_B}\sqrt{\frac{\left(n-\frac{m+|m|}{2}\right)!}{\left(n+\frac{|m|-m}{2}\right)!}}, 
\end{eqnarray}
where $l_B=\frac{1}{\sqrt{eB}}$ is called the magnetic length, $L^{m}_{n}(z)$ is the $n$-th order associated Laguerre polynomial. We recall that this wave function is also the eigen state of the canonical OAM, $L^{\rm can}_{z} = -i\frac{\partial}{\partial \phi}$, i.e. $\hat{L}^{\rm can}_{z}\psi^{(\rm S)}_{n,m} = m \psi^{(\rm S)}_{n,m}$ with $m \le n$.

\subsection{Gauge invariant pseudo OAM and its observability}
Using the canonical momentum or the mechanical momentum, one can construct two OAMs, that is, the canonical OAM, $\hat{\textbf L}^{\rm can} = {\textbf r} \times \hat{\textbf p}$, and the mechanical OAM, $\hat{\textbf L}^{\rm mech} = {\textbf r} \times \hat{\bm \Pi}$, where the canonical OAM is not gauge-invariant, while the mechanical one is invariant under the local gauge transformation. It is usually believed that the physical OAM is the mechanical one according to the gauge principle and classical correspondence \cite{Murayama}. However, we have another gauge-invariant OAM in this system, i.e. the pseudo OAM which is recently discussed \cite{KM1,KM2}, 
\begin{eqnarray}
 \hat{L}^{\rm ps}_{z} \equiv \hat{L}^{\rm mech}_{z} - \frac{eB}{2}r^2.
\end{eqnarray}
This OAM commutes with the Hamiltonian, $[\hat{L}^{\rm ps}_{z},H]=0$, in arbitrary gauges and it reduces to the canonical OAM in the symmetric gauge. Hence this gives an example of the so-called gauge-invariant-extension of the canonical OAM \cite{Nspin_review1}.
 Actually, we notice that this pseudo-OAM was already discussed by Johnson and Lippmann many years ago for a different purpose \cite{JohnsonLippmann1,JohnsonLippmann2}. They used this conserved operator and its eigen equation for the ground state to obtain wave functions for a higher Landau-level.

Our previous study \cite{path_dep_Wakamatsu} based on the DeWitt's gauge-invariant formulation of the electrodynamics gives the following expectation values for the canonical, mechanical, and pseudo OAMs:
\begin{eqnarray}
\langle \hat{L}^{\rm can}_{z} \rangle 
&=& m, \hspace{1cm}
\langle \hat{L}^{\rm mech}_{z}\rangle = 2n+1, \hspace{1cm} \langle \hat{L}^{\rm ps}_{z} \rangle  = m, \label{eq.quantum.Lcan.Lmech.Lps.origin}
\end{eqnarray}
where $m$ is the eigen value of the canonical and pseudo OAM with respect to the origin. The readers who are not familiar with the gauge-invariant method can consult with section 3 in our previous work \cite{path_dep_Wakamatsu} for the technical detail. 

Although the above results show that both the mechanical OAM $\langle \hat{L}^{\rm mech}_z \rangle$ and the pseudo OAM $\langle \hat{L}^{\rm ps}_z \rangle$ seem to have the qualification to be an observable because of its gauge-invariant nature, we concluded in our previous work \cite{path_dep_Wakamatsu} that the physical OAM corresponding to an observable is the mechanical OAM $\langle \hat{L}^{\rm mech}_{z} \rangle$, not pseudo OAM $\langle \hat{L}^{\rm ps}_{z} \rangle$. This is simply because that the mechanical OAM is proportional to the Landau energy which is definitely the observable of this system. However, the relation between the gauge invariance and the observability or non-observability of the gauge-invariant-canonical OAM, is not clear enough. In the next section, we analyze this question by a different viewpoint based on the relation between the guiding center and reference axes for OAMs in this system.

\section{Guiding center in classical theory and quantum theory \label{sec3}}
In this section, we consider the role of the guiding center and OAMs for the electron moving in the uniform magnetic field both in classical theory and quantum theory. First, we begin our discussion of the guiding center and the mechanical and pseudo OAMs with the well-known cyclotron motion in classical theory. Next, we clarify the role of guiding center and the above two OAMs in quantum theory by using the analogy with classical picture.

\subsection{OAMs and guiding center in classical theory}
The motion of the electron with the charge $-e~(e>0)$ and the mass $m_e$ in the classical theory is determined by equation of motion (EOM):
\begin{eqnarray}
 m_e\dot{{\textbf v}}(t) = -e({\textbf v}(t)\times {\textbf B}), \label{eq.EOM.Lorentz}
\end{eqnarray}
where the dot means the time derivative and ${\textbf v}(t)\equiv \dot{{\textbf x}}(t)$. As is well-known, the general solutions for the electron's orbit $\left(x(t),y(t)\right)$ in the two-dimension plane are given by,
\begin{eqnarray}
 x(t) &=& X + \frac{1}{\omega}v_{y}(t), \hspace{1cm} X = x_0 - \frac{v_{x0}}{\omega}, \nn\\
 y(t) &=& Y - \frac{1}{\omega}v_{x}(t), \hspace{1cm} Y = y_0 - \frac{v_{y0}}{\omega}, 
\end{eqnarray}
where $v_{x}(t) = v_{0}\cos(\omega t + \alpha),~v_{y}(t) = v_{0}\sin(\omega t + \alpha),~v_{0} = \sqrt{v^2_{x0} + v^2_{y0}},~\tan\alpha = \frac{v_{y0}}{v_{x0}}$. The solutions satisfy the initial conditions, $x(0) = x_{0}, v_{x}(0) = v_{x0}, y(0)  = y_{0}, v_{y}(0) = v_{y0}$, and $\omega =\frac{eB}{m_e}$ is the cyclotron frequency. In the classical mechanics, the coordinate $(X,Y)$ is called the guiding center which has the clear physical meaning as the center of the cyclotron motion. Importantly, this guiding center $(X,Y)$ is time independent, $\dot{X}=\dot{Y}=0$. 

In addition to these conserved quantities, $X$ and $Y$,  we can find the conserved angular momentum called the pseudo OAM \cite{ KM1,KM2} from EOM in (\ref{eq.EOM.Lorentz}),
\begin{eqnarray}
 \frac{d}{dt}L^{\rm ps}_{z} &\equiv& \frac{d}{dt}\left[ {\textbf x}(t)\times m_e{\textbf v}(t) - \frac{e}{2}r^2(t){\textbf B}\right]_{z}=0. \label{eq.classical.def.pseudo.OAM}
\end{eqnarray}
Note, however, that this conserved OAM, $L^{\rm ps}_{z}$, does not correspond to the well known electron's classical OAM. For later discussion in quantum theory, we consider the two cases, $(X,Y)=(0,0)$ and $(X,Y)\neq (0,0)$, separately. 

We first point out that there is one subtlety in discussing OAM, namely, we must necessarily specify a reference axis for OAM and this point is different from the case of linear momenta. With this point in mind, we first compare the mechanical OAM and the pseudo OAM in classical theory.

We consider two reference-axes for the mechanical OAM and the pseudo OAM, namely, the origin and the guiding center which is different from the origin. Hence we define the following four OAMs in total:
\begin{eqnarray}
 L^{\rm \ mech}_z &\equiv& \left[{\textbf x}(t)\times m_e{\textbf v}(t)\right]_z, \label{eq.classical.def.Lmech.origin}\\
 L^{\rm ps}_z &\equiv& \left[{\textbf x}(t)\times m_e{\textbf v}(t) - \frac{e}{2}r^2(t){\textbf B}\right]_z,\label{eq.classical.def.Lps.origin}\\
  \mathcal{L}^{\rm  mech}_z &\equiv& \left[\left({\textbf x}(t)-{\textbf R} \right)\times m_e{\textbf v}(t)\right]_z, \label{eq.classical.def.Lmech.GC}\\
 \mathcal{L}^{\rm ps}_z &\equiv& \left[\left({\textbf x}(t)-{\textbf R}\right)\times m_e{\textbf v}(t) - \frac{e}{2}\left({\textbf x}(t)-{\textbf R}\right)^2{\textbf B}\right]_z, \label{eq.classical.def.Lps.GC}
\end{eqnarray}
where ${\textbf R}=(X,Y)$ is the relative position vector to the guiding center from the origin and ${\textbf x}(t)=(x(t),y(t))$ is the coordinate vector to the electron's position from the origin. The first two OAMs~($L^{\rm mech}_z,L^{\rm ps}_{z}$) express the mechanical and the pseudo OAM with respect to the origin, while the last two OAMs~($\mathcal{L}^{\rm mech}_{z}, \mathcal{L}^{\rm ps}_z$) express the mechanical and pseudo OAM with respect to the guiding center ${\textbf R}$.

\subsubsection{OAMs in the case A: $(X,Y)=(0,0)$}
First, we set $(X,Y)=(0,0)$. This choice of the initial conditions corresponds to the case (A) of Fig.~\ref{fig.guiding.center.cl}. After one substitutes the solutions of the EOM into each definition, the mechanical and pseudo OAMs for $(X,Y)=(0,0)$ are reduced to
\begin{eqnarray}
 L^{\rm  mech}_{z} &=& \mathcal{L}^{\rm mech}_{z} 
 =  r_c m_ev_{0}, \label{eq.def.mech.0}\\
 L^{\rm ps}_{z} 
 &=& \mathcal{L}^{\rm ps}_{z}
  =  \frac{r_c m_e v_{0}}{2}, \label{eq.def.KM.0}
\end{eqnarray}
where $r_{c} = \frac{v_0}{\omega}$ is the cyclotron radius.
Here, it is obvious that $L^{\rm mech}_{z}$ gives the well-known conserved and classical OAM, $r_c m_ev_{0}$, of the cyclotron motion with respect to the origin. On the other hand, we find that $L^{\rm ps}_{z}$ gives just one half of the mechanical OAM.  

\subsubsection{OAMs in the case B: $(X,Y) \neq (0,0)$}
\begin{center}
 \begin{figure}[t]
  \def\SCALEA{0.40}
  \def\SCALEB{0.38}
  \def\OFFSET{10pt}
  \begin{tabular}{cc}
   \includegraphics[scale=\SCALEA]{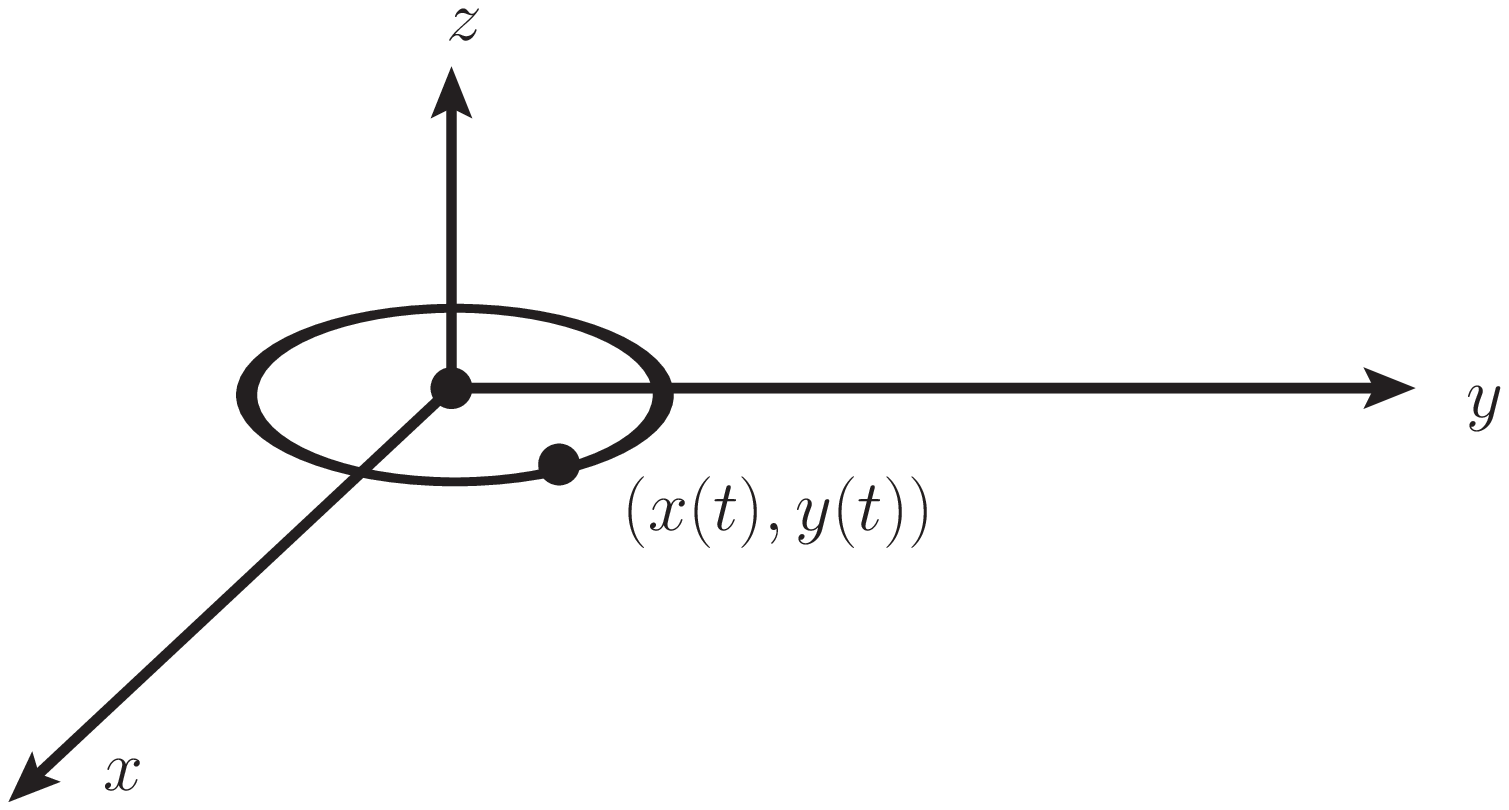} &
   \includegraphics[scale=\SCALEB]{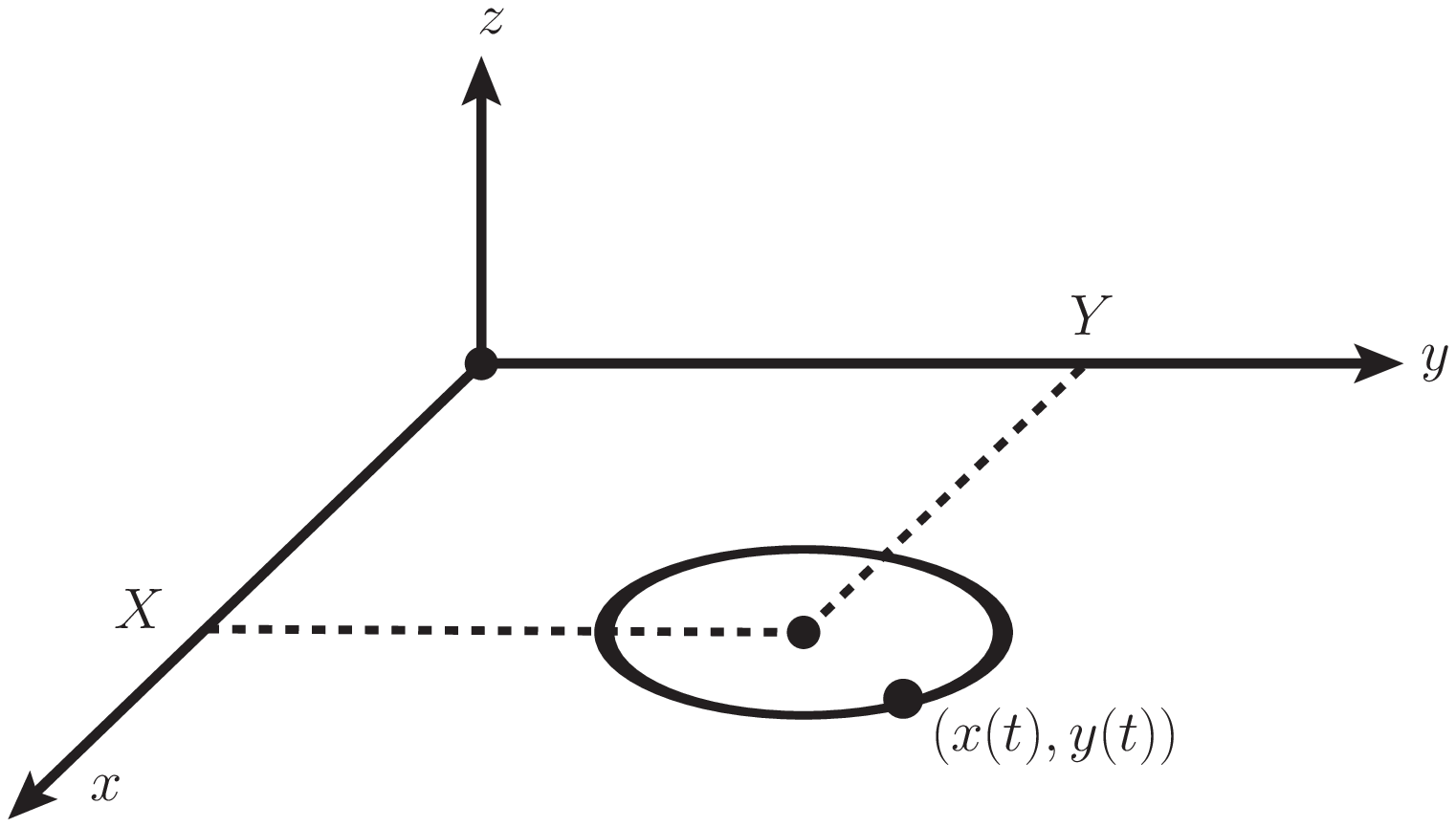} \\
   (A) & (B)
  \end{tabular}
\caption{The cyclotron motion of the electron at $z=0$ plane in classical mechanics: the case (A) for $(X,Y)=(0,0)$ and the case (B) for $(X,Y)\neq(0,0)$. The orbit of the electron is expressed by $(x(t),y(t))$ as the function of time. } \label{fig.guiding.center.cl}
 \end{figure}
\end{center}

Next, we set $(X,Y) \neq (0,0)$. This choice corresponds to the case (B) of Fig.~\ref{fig.guiding.center.cl}. We consider the two sets of OAMs, namely, $(L^{\rm  mech}_{z}, L^{\rm ps}_{z})$ defined in (\ref{eq.classical.def.Lmech.origin}),(\ref{eq.classical.def.Lps.origin}), and  $(\mathcal{L}^{\rm  mech}_{z}, \mathcal{L}^{\rm ps}_{z})$ defined in (\ref{eq.classical.def.Lmech.GC}),(\ref{eq.classical.def.Lps.GC}), for $(X,Y) \neq (0,0)$. These four OAMs reduce to:
\begin{eqnarray}
 L^{\rm  mech}_{z} 
 &=& r_c m_e v_0 + m_e\left[ Xv_{y}(t) - Y v_{x}(t)\right],\label{eq.classical.result.Lmech.origin.caseB}\\
 L^{\rm ps}_{z} 
 &=& \frac{1}{2}r_c m_e v_{0} - \frac{eB}{2}\left[X^2+Y^2\right],\label{eq.classical.result.Lps.origin.caseB}\\
 \mathcal{L}^{\rm  mech}_{z} 
 &=& r_c m_e v_{0},\label{eq.classical.result.Lmech.GC.caseB}\\
 \mathcal{L}^{\rm ps}_{z}
 &=& \frac{1}{2}r_c m_e v_{0}.\label{eq.classical.result.Lps.GC.caseB}
\end{eqnarray} 
We find that $L^{\rm mech}_{z}$ consists of two terms, i.e. the first time-independent term and the second time-dependent term. The first term just coincides with the well-known angular momentum corresponding to the classical cyclotron motion, while the second term vanishes if we take the time-average over one period of the cyclotron motion. Namely, we have
\begin{eqnarray}
 \langle L^{\rm mech}_{z} \rangle_{T}
 \equiv \frac{1}{T} \int_{0}^{T} dt L^{\rm mech}_{z} 
 = r_{c} m_e v_0,
\end{eqnarray}
where $T\equiv2\pi/\omega$.

$L^{\rm ps}_{z}$ also consists of two terms. The first term is just one half of the OAM corresponding to the cyclotron motion, while the second term is a function of the square of the distance from the origin to the guiding center and hence it depends on a coordinate choice. Different from the above two OAMs defined with respect to the origin, the corresponding two OAMs $\mathcal{L}^{\rm mech}_{z}$ and $\mathcal{L}^{\rm ps}_{z}$ defined with respect to the guiding center turn out to be both time-independent and also independent of the guiding-center-coordinates $X$ and $Y$. As one sees, $\mathcal{L}^{\rm mech}_{z}$ coincides with the well-known OAM corresponding to the classical cyclotron-motion, whereas $\mathcal{L}^{\rm ps}_{z}$ is just one half of it.    

\subsection{OAMs and guiding center in quantum theory}
In quantum theory, the mechanical momentum is replaced by the operator, $\hat{\bm \Pi}=-i{\bm \nabla} + e{\textbf A}$, and the guiding center is also replaced by the operator:
\begin{eqnarray}
 \hat{X} = x - \frac{1}{eB}\hat{\Pi}_{y}, 
\hspace{1cm}
 \hat{Y} = y + \frac{1}{eB}\hat{\Pi}_{x}.
\end{eqnarray} 
It is important to recognize that the guiding center in quantum theory is a $q$-number and we cannot freely set $(\hat{X},\hat{Y})$ zero, like a $c$-number in classical theory. In this sense, the meaning of the guiding-center-operator in the quantum theory is far less intuitive.

Very interestingly, the guiding-center-operator $(\hat{X},\hat{Y})$ is time-independent even in quantum theory, namely, we have $\left[\hat{X},\hat{H}\right] = \left[\hat{Y},\hat{H}\right] = \left[\hat{R}^2, \hat{H}\right] = 0$, where $\hat{H}$ stands for the quantized Landau-Hamiltonian and $\hat{R}^2 \equiv \hat{X}^2+\hat{Y}^2$. 
However, these two coordinate operators do not commute with each other, and satisfy the commutation relation, $\left[\hat{X},\hat{Y}\right] = il^2_{B}$.
The last commutation-relation indicates the Heisenberg's uncertainty relation between $\hat{X}$ and $\hat{Y}$, that is, we cannot precisely specify the position of $\hat{X}$ and $\hat{Y}$ simultaneously. 

Thus we now encounter a very peculiar situation. In quantum theory the guiding center still has a meaning as a center of cyclotron motion, but here is an inherit uncertainty in its position. This motivates us to consider OAMs defined with respect to the quantum guiding-center. We define four types of OAMs:  
\begin{eqnarray}
\hat{L}^{\rm  mech}_{z} &\equiv& \left[ {\textbf x} \times \hat{\bm \Pi}\right]_{z}, \label{eq.quantum.Lmech.origin}\\
 \hat{L}^{\rm ps}_{z} &\equiv& \left[{\textbf x} \times \hat{\bm \Pi} - \frac{e}{2}\left(x^2+y^2\right){\textbf B} \right]_{z}, \label{eq.quantum.Lps.origin}\\
 \hat{\mathcal{L}}^{\rm  mech}_{z} 
 &\equiv& \left[\left({\textbf x}-\hat{\textbf R} \right)\times \hat{{\bm \Pi}} \right]_z, \label{eq.quantum.Lmech.GC}\\
 \hat{\mathcal{L}}^{\rm ps}_{z} 
 &\equiv& \left[\left({\textbf x}-\hat{\textbf R}\right)\times \hat{\bm \Pi} - \frac{e}{2}\left({\textbf x}-\hat{\textbf R}\right)^2{\textbf B}\right]_z, \label{eq.quantum.Lps.GC}
\end{eqnarray}
where ${\textbf x}=(x,y)$ and $\hat{\textbf R}=(\hat{X},\hat{Y})$ are the coordinate for the electron's position and the quantum guiding-center-operator, respectively.
The first two OAMs, $(\hat{L}^{\rm  mech}_{z},\hat{L}^{\rm  ps}_{z})$, are the quantum mechanical-OAM and pseudo-OAM with respect to the origin. On the other hand, the last two OAMs, $(\hat{\mathcal{L}}^{\rm  mech}_{z}, \hat{\mathcal{L}}^{\rm  ps}_{z})$, are the quantum mechanical-OAM and pseudo-OAM with respect to the quantum guiding-center. (The OAM around the guiding center was also investigated in a recent paper by van Enk \cite{van_Enk}.) In addition, in quantum theory, we also consider the canonical OAMs given by:
\begin{eqnarray}
\hat{L}^{\rm  can}_{z} &\equiv& \left[ {\textbf x} \times \hat{\textbf p}\right]_{z},\label{eq.quantum.Lcano.origin}\\
\hat{\mathcal{L}}^{\rm  can}_{z} &\equiv& \left[ \left({\textbf x} - \hat{\textbf R} \right) \times \hat{\textbf p}\right]_{z}, \label{eq.quantum.Lcano.GC}
\end{eqnarray}
where $\hat{L}^{\rm can}_{z}, \hat{\mathcal{L}}^{\rm can}_{z}$ stand for the canonical OAM with respect to the origin and with respect to the quantum guiding-center, respectively. 

As pointed out above, a delicate point here is that the guiding center in quantum theory is a q-number operator and there is an inherent uncertainty in its position (see the figures 1,2 in the reference \cite{CHLi_QWang}, and the figure~7 in the reference \cite{VGW}). Still, we need to specify the coordinate system, in particular, the origin, for solving the Schr\"{o}dinger equation. Then, the standard procedure is first to choose the origin of the coordinate arbitrarily, i.e. on no account of the guiding-center concept. After that, we choose a gauge potential which reproduces the uniform magnetic field. The most convenient choice for our discussion is the use of the symmetric-gauge potential and the associated wave-functions. 

First, we consider three OAMs, namely, the mechanical, the pseudo, and the canonical OAMs with respect to the origin. Their expectation values are already given in (\ref{eq.quantum.Lcan.Lmech.Lps.origin}), so that we do not show them again. 

Next, we consider the three OAMs $\hat{\mathcal{L}}^{\rm can}_{z}$, $\hat{\mathcal{L}}^{\rm mech}_{z}$, and $\hat{\mathcal{L}}^{\rm ps}_{z}$, which are defined with respect to the guiding center. First, by using the relation $\hat{\textbf{p}}=\hat{\boldsymbol {\Pi}}-e\textbf{A}$, the canonical OAM  $\hat{\mathcal{L}}^{\rm can}_{z}$ is written as 
\begin{eqnarray}
  \hat{\mathcal{L}}^{\rm can}_{z} =  \hat{\mathcal{L}}^{\rm mech}_{z} - e\left[ (\textbf{x}-\hat{\textbf{R}})\times \textbf{A} \right]_{z}.
\end{eqnarray}
On the other hand, by using the relation,
\begin{eqnarray}
 (\textbf{x}-\hat{\textbf{R}})\times\boldsymbol{\Pi} = \frac{1}{eB}\boldsymbol{\Pi}^2,
\end{eqnarray}
together with $\omega=eB/m_e$, the mechanical OAM $\hat{\mathcal{L}}^{\rm mech}_{z}$ can be expressed as
\begin{eqnarray}
 \hat{\mathcal{L}}^{\rm mech}_{z} = \frac{2}{\omega}\hat{H}.
\end{eqnarray}
Finally, the first term of $\hat{\mathcal{L}}^{\rm ps}_{z}$ given in (\ref{eq.quantum.Lps.GC}) equals to $\hat{\mathcal{L}}^{\rm mech}_{z}=2\hat{H}/\omega$. On the other hand, owing to the relation $(\textbf{x}-\hat{\textbf{R}})^2=\hat{\boldsymbol{\Pi}}^2/(eB)^2$, the second term in (\ref{eq.quantum.Lps.GC}) equals to $-\hat{H}/\omega$. It cancels a half of the first term, thereby leading to
\begin{eqnarray}
 \hat{L}^{\rm ps}_{z} = \frac{1}{\omega}\hat{H}.
\end{eqnarray}
Summarizing the results for the three OAMs $\hat{\mathcal{L}}^{\rm can}_{z}$, $\hat{\mathcal{L}}^{\rm mech}_{z}$, and $\hat{\mathcal{L}}^{\rm ps}_{z}$, we thus obtain
\begin{eqnarray}
 \hat{\mathcal{L}}^{\rm  can}_{z}
  &=& \frac{2}{\omega}\hat{H}-\frac{1}{2}\hat{L}^{\rm  can}_{z} - \frac{eB}{4}r^2, \hspace{0.5cm} 
 \hat{\mathcal{L}}^{\rm  mech}_{z} 
  = \frac{2}{\omega}\hat{H}, \hspace{0.5cm}
 \hat{\mathcal{L}}^{\rm ps}_{z} 
 = \frac{1}{\omega} \hat{H}. \label{eq.Lmech.Lps.operator.relation.GC}
\end{eqnarray}
Very interestingly, $\hat{\mathcal{L}}^{\rm  mech}_{z}$ and $\hat{\mathcal{L}}^{\rm  ps}_{z}$ are both proportional to the Landau Hamiltonian and hence the conservation of these expectation values are self-evident. 

Now it is an easy exercise to evaluate the expectation values of these three OAM operators between the Landau eigen-states in the symmetric gauge. The answers are given by:

\begin{eqnarray}
\langle \hat{\mathcal{L}}^{\rm  can}_{z} \rangle &=& \frac{2n+1}{2}, \hspace{1cm}
\langle \hat{\mathcal{L}}^{\rm  mech}_{z} \rangle
= 2n+1, \hspace{1cm}
\langle \hat{\mathcal{L}}^{\rm  ps}_{z} \rangle
= \frac{2n+1}{2}. \label{eq.quantum.Lmech.Lps.GC}
\end{eqnarray}
 
For readers who are not very familiar with the Landau problem, we think it instructive to add some explanation, which is expected to clarify the reason why the mechanical OAM of the electron takes only odd values (in the unit of $\hbar$). In view of the proportionality relation between the Landau Hamiltonian and the mechanical OAM defined with respect to the guiding center, the question amounts to asking why the Landau energies take only odd values in unit of Larmor frequency $\omega_L$ as $E=(2n+1)\omega_L$. Note that, if this same energy is expressed in terms of the cyclotron frequency $\omega=2\omega_L$, it takes the form $E=(n+1/2)\omega$. Hence, it seems to be a matter of convenience whether one says the Landau energy is proportional to odd integer or it is proportional to half integer. Incidentally, it is a widely known fact that the
Landau Hamiltonian is expressed as a sum of two-dimensional Harmonic oscillator Hamiltonian
$H_{\rm osc}$ and the Larmor energy term $H_{\rm Larmor} = \omega_{L} L^{\rm can}_{z}$. The expectation value of the Larmor term in the Landau state $|n,m\rangle$ is given by $\langle H \rangle = m\omega_L$, while the expectation value of the two-dimensional Harmonic oscillator is given by $\langle H_{\rm osc} \rangle = (2n-m+1) \omega_L$, which add up to
give the total energy $(2n+1) \omega_L$. We recall that, in the unit of $\omega_L$, the term of unity in
the energy of the two-dimensional Harmonic oscillator comes from the zero-point energy of the two-dimensional Harmonic oscillator. It is unity, because it is the sum of the zero-point energy $1/2$ in each of the two dimensions.

Another more intuitive explanation may be obtained by a semi-classical argument. Based on the familiar Bohr-Sommerfeld semi-classical quantization rule for the closed cyclotron orbit:
\begin{eqnarray}
 n + \frac{1}{2} = \oint {\mathbf p}\cdot d{\mathbf r}.
\end{eqnarray}
Here, the right-hand-side represents a closed line integral of the canonical momentum operator ${\mathbf p}$. From this, one obtains 
\begin{eqnarray}
 \left( n + \frac{1}{2} \right)\omega = \frac{m_e}{2}v^2 = E.
\end{eqnarray}  
Combining this relation with $\omega=v/r$, we immediately obtain 
\begin{eqnarray}
L^{\rm mech}_{z} = m_e r v = 2n+1,
\end{eqnarray}
which explains why the mechanical OAM of the electron takes only odd values. 

Still another illuminating explanation as indicated from the paper \cite{Greenshields2} would be following. The mechanical OAM $\hat{\mathcal{L}}^{\rm mech}_{z}$ defined with respect to the guiding center is associated with the rotation around the guiding center at the cyclotron frequency, whereas the diamagnetic OAM, i.e. the second term of Eq.~(\ref{eq.quantum.Lps.GC}), is associated with the rotation about the center-of-mass of the wavepacket at the Larmor frequency, which is half of the cyclotron frequency. This means that half of the 1st term of Eq.~(\ref{eq.quantum.Lps.GC}) is cancelled by the second term, thereby explaining the factor $1/2$ difference between the expectation values of $\hat{\mathcal{L}}^{\rm ps}_{z}$.

\section{Discussion \label{sec4}}
A unique (or singular) feature of the Landau problem is that the magnetic field is uniformly spread over the whole two-dimensional plane, so that there is no special point in this plane. In classical theory, the center of electron's cyclotron motion is uniquely fixed by the initial conditions. Though the choice of the coordinate system for describing the Landau system is totally arbitrary, the most convenient choice of it would be to choose the center of the cyclotron motion as the coordinate origin. In quantum mechanics, however, the center of the cyclotron motion, also called the guiding center, is a q-number quantity so that there is an inherent uncertainty in its position even after appropriately fixing the coordinate system. This motivates us to consider two types of OAM, i.e. the OAM around the coordinate origin and that around the quantum guiding-center. Besides, since our original motivation is to understand the difference of the three OAMs, i.e. the canonical OAM, the mechanical OAM, and the pseudo OAM, we consider totally six OAMs. They are the canonical OAM $\hat{L}^{\rm can}_{z}$, the mechanical OAM $\hat{L}^{\rm mech}_{z}$, and the pseudo OAM $\hat{L}^{\rm ps}_{z}$ defined with respect to the coordinate origin, and the canonical OAM $\hat{\mathcal{L}}^{\rm can}_{z}$, the mechanical OAM $\hat{\mathcal{L}}^{\rm mech}_{z}$, and the pseudo OAM $\hat{\mathcal{L}}^{\rm ps}_{z}$ defined with respect to the quantum guiding-center.
We summarize the expectation values of these OAMs in the Landau eigen-states in Table \ref{table1}.

\begin{table}[ph]
 \tbl{Summary of the expectation values for the three OAMs defined with respect to the origin and guiding center.}
{\begin{tabular}{@{}cccc@{}} \toprule
     Axis of OAM &                                                 Canonical OAM &                                       Mechanical OAM &                                                  Pseudo OAM\\ 
\colrule
          Origin & $\langle \hat{L}^{\rm can}_z \rangle = m$                        &           $\langle \hat{L}^{\rm mech}_z \rangle = 2n+1$ &                       $\langle \hat{L}^{\rm ps}_z \rangle = m$\\ 
\colrule
  Guiding center & $\langle \hat{\mathcal{L}}^{\rm can}_z \rangle = \frac{2n+1}{2}$ & $\langle \hat{\mathcal{L}}^{\rm mech}_z \rangle = 2n+1$ & $\langle \hat{\mathcal{L}}^{\rm ps}_z \rangle = \frac{2n+1}{2}$\\
\botrule
   \end{tabular} \label{table1}}
\end{table}
First, we point out that the expectation values of $\hat{L}^{\rm mech}_{z}$ and $\hat{\mathcal{L}}^{\rm mech}_{z}$ exactly coincide with each other in perfect conformity with the classical consideration given in section \ref{sec3}, which gives the equality $\langle L^{\rm mech}_{z} \rangle_{T} = \langle \mathcal{L}^{\rm mech}_{z} \rangle_{T}$ for the time-averaged two OAMs. These expectation values are expressed with the so-called Landau quantum-number $n$ characterizing the eigen-energies of the Landau level system.  This is only natural because the quantized Landau energy is the energy of the cyclotron motion with respect to some point which is arbitrary in the specific case of the Landau problem. It is again interpreted as showing the physical nature of the mechanical OAM. 

Also noteworthy are the relations, $\langle \hat{L}^{\rm can}_z \rangle = \langle \hat{L}^{\rm ps}_z\rangle$ and $\langle \hat{\mathcal{L}}^{\rm can}_z\rangle = \langle \hat{\mathcal{L}}^{\rm ps}_z\rangle$. These are in some sense expected relations, because the pseudo OAMs reduce to the canonical OAMs in the symmetric gauge aside from the unphysical gauge degree of freedom. From the physical viewpoint, then, the pseudo OAM and canonical OAM are essentially the same quantity. This strongly indicates that the pseudo OAM, just like the ordinary canonical OAM, need not correspond to an observable in spite of its formal gauge-invariance. 

Another non-trivial observation here is the relations, $\langle \hat{L}^{\rm ps}_{z} \rangle \neq \langle \hat{\mathcal{L}}^{\rm ps}_{z} \rangle$ and $\langle \hat{L}^{\rm can}_{z} \rangle \neq \langle \hat{\mathcal{L}}^{\rm can}_{z} \rangle$. It reconfirms us that the concept of the pseudo OAM or canonical OAM vitally depends on the choice of the coordinate system. The expectation values of the pseudo OAM and the canonical OAM with respect to the origin both depend on the quantum number $m$. This quantum number $m$ is a conserved quantity reflecting the rotational (or axial) symmetry of the Hamiltonian. This symmetry is not independent of the choice of the coordinate origin and the vector potential, both of which have large arbitrariness, because of the perfect uniformity of the magnetic field in the Landau level system. The dependence on the choice of the coordinate origin must be the source of non-observability of the quantum number $m$ in the Landau-level system. 

On the contrary, the expectation value of the pseudo OAM and/or the canonical OAM defined with respect to the guiding center turns out to be expressed with the Landau quantum-number $n$, which means that they are related to an observable. Note that, since these OAMs are defined with respect to the guiding center, which can be interpreted as the center of the cyclotron motion, they have definite physical meanings, which are independent of the choice of the coordinate origin. Undoubtedly, this is the reason why they are related to the observable
through the Landau quantum-number $n$.

Here, to avoid possible confusion or misunderstanding, we want to point out a delicate difference between the guiding center concept and the "off-axis centroid" discussed in the paper \cite{Greenshields2,Schachinger}. The "off-axis centroid" is somehow a semi-classical concept. In contrast, the concept of the guiding center in the Landau problem is a genuine quantum mechanical one. As discussed in our recent papers \cite{Wakamatsu_PLA}, in quantum mechanics, there is an unavoidable quantum mechanical uncertainty in the position of the guiding center.  When the eigen-state $|n,m\rangle$ of the Landau Hamiltonian is given, the guiding center is distributed on a circle of the radius $R$ with equal probability. Here the radius $R$ represents the average distance between the guiding center and the coordinate origin. As a consequence, the centroid of the Landau electron is still located on the coordinate origin not on the centroid of the each cyclotron motion. Although all these features were already explained in our recent paper \cite{Wakamatsu_PLA}, it may be instructive to remember the important identity, which relates the canonical OAM $\hat{L}^{\rm can}_{z}$ and two radii, i.e. the cyclotron radius $\hat{r}_c$ and the distance $\hat{R}$ between the guiding center and the coordinate origin. This important relation, which was first written down by Johnson and Lippmann many years ago\cite{JohnsonLippmann1,JohnsonLippmann2}, reads as
\begin{eqnarray}
 \hat{L}^{\rm can}_{z} = \frac{1}{2l^2_B}\left(\hat{r}^2_c - \hat{R}^2\right), 
\end{eqnarray}
with $l_B=1/\sqrt{eB}$ being the familiar magnetic length in the Landau problem. Taking the expectation value of the above equation, one finds that 
\begin{eqnarray}
r^2_c &\equiv& \langle \hat{r}^2_c \rangle = (2n+1)l^2_B,  \\
 R^2 &\equiv& \langle \hat{R}^2 \rangle = (2n-2m+1)l^2_B,
\end{eqnarray}
which in turn gives
\begin{eqnarray}
 \langle \hat{L}^{\rm can}_{z} \rangle = m,
\end{eqnarray}
as it should be. A remarkable consequence of the Johnson and Lippmann's relation is that the sign of the quantum number $m$ is inseparably connected with the magnitude correlation between the two radii $r_c$ and $R$. (See \cite{Wakamatsu_PLA}, for more detail.)

We have argued that the magnetic quantum number $m$ of the Landau problem does not correspond to observables and that this is not inconsistent with the gauge-variant nature of the canonical OAM. On the other hand, it is a widely-known fact that the magnetic quantum number $m$ of the helical or twisted electron beam like the Laguerre-Gauss beam (this quantum number is often called the topological index of the twisted beam) is in fact an observable quantity. To avoid a misunderstanding, we recall the fact that the standard Laguerre-Gauss beam is an approximate solution of the free wave equation of the electron. In the free space, there is no difference between the {\it gauge-variant} canonical OAM and the {\it gauge-invariant} mechanical or kinetic OAM. This is because the gauge potential part simply vanishes for a free electron. In the general case of non-free electron, however, what appears in the equation of motion of the electron is the gauge-invariant mechanical momentum and mechanical OAM not the canonical ones. Accordingly, the change or transfer of the electron's mechanical OAM can in principle be determined by measuring the torque which the electron beam exerts on some external atom or something. This mechanical OAM is generally made up of the canonical OAM and the gauge potential part. Since these two OAMs always appear as a single combination in the equation of motion, each cannot be separately measured at least by means of torque measurements. For free electron beam, however, there is no gauge potential part, so that the canonical part directly enters the torque equation. As a consequence, observation of the canonical OAM for a free electron beam does not contradict the gauge principle, which dictates that observables must be gauge-invariant. Exactly the same can be said also for the observation of the OAM of photon in laser physics.

Still highly nontrivial is the observation made by Bliohk et al. \cite{Bliokh_PRX} in the physics of helical electron beam propagating along a uniform magnetic field. In this paper, they notice the similarity between the Landau eigenfunctions and the wave functions of the transverse part of the Laguerre-Gauss beam. They argue that, allowing free propagation along the direction of a uniform magnetic field, the familiar Landau electron state can be regarded as a non-diffracting version of the helical electron beam propagating along the direction of the magnetic field. Starting with this observation, they showed that, while propagating along the magnetic field, the Landau electrons receive characteristic rotation with three different angular velocities, depending on the eigen-value $m$ of the canonical OAM operator, and this splitting was later experimentally confirmed in the paper \cite{Lcan_Nature}.

At first sight, this appears to provide evidence for the observation of the quantum number $m$ of the Landau states. Since the Landau electron is not a free particle, the observation of the eigenvalue of the gauge-variant canonical OAM in this case might contradict the gauge principle. Then, does this observation give an evidence that the gauge-principle is not always correct ?

In our recent paper \cite{Wakamatsu_PLA}, we have carefully examined this delicate problem. We showed there that the above-mentioned $m$-dependent splitting of the electron's rotational trajectory can be explained only on the basis of the gauge-invariant mechanical (or total) current density, so that it does not necessarily contradict the gauge principle. We also showed that the 3-fold splitting of the electron's rotational trajectory depending on the sign of $m$ has a simple and intuitive explanation based on the quantum guiding center concept in the Landau problem. Namely, the novel $m$-dependent splitting of the electron's rotational motion, while propagating along the magnetic field, can be understood if one notices the magnitude correlation between the cyclotron radius and the distance of the guiding center from the coordinate origin, which critically depends on the value of $m$. Finally, we also emphasized the fact that not all the degeneracies of the Landau states on the quantum number $m$ are resolved by their experimental setup by using the helical electron beam. In our conjecture, still remaining degeneracy of the rotational frequency for both of the $m > 0$ mode and of the $m < 0$ mode, should rather be interpreted as a consequence of the gauge-invariant requirement for observables.

At the present moment, we are not completely sure whether one can find still cleverer setup for observing the magnetic quantum number $m$ or the canonical OAM in the presence of non-zero background of electromagnetic field not in a free space. It is still an interesting open question.

\section{Conclusion \label{sec5}}
The gauge principle dictates that observables must be gauge invariant. There is no doubt about the validity of the gauge principle, especially for some simple observables like the S-matrix. For some delicate problems, however, the relation between the gauge-invariance and the observability is far less clear than one would naively think. In the present paper, we have intendedly compared two formally gauge-invariant OAMs in the Landau problem, i.e. the pseudo OAM defined with respect to the origin and the pseudo OAM defined with respect to the guiding center. We showed that, although these OAMs are both formally gauge invariant, only the latter can be related to an observable. The reason of non-observability of the former quantity seems to be intimately connected with the special nature of the Landau problem, in which the magnetic field spreads uniformly over the whole two-dimensional plane, so that there is no particular point in this plane. The magnetic quantum number $m$ as the eigenvalue of the canonical OAM operator is certainly an important (conserved) quantum number reflecting the axial symmetry of the Landau Hamiltonian. However, this symmetry of the Landau Hamiltonian is critically dependent on the choice of the coordinate origin as well as the choice of the symmetric gauge. This arbitrariness is the ultimate origin of the infinite degeneracy of the Landau states with respect to the quantum number $m$, thereby leading to the non-observability of the canonical OAM (and/or the pseudo OAM) in the Landau problem.

It should be contrasted with the pseudo OAM defined with respect to the guiding center, which has a clear physical meaning independently of the choice of the coordinate system or the coordinate origin. To sum up, the so-called gauge principle claims that observables must be gauge invariant, and this fundamental principle of physics is widely believed to be correct. However, we must be careful about the fact that the converse of this theorem is not necessarily true. As we have shown through several concrete examples, the formal gauge invariance of some quantities does not always ensures its observability. In such an occasion, the gauge symmetry possessed by those quantities are thought of as just a formal redundancy without little physical significance.

\section*{Acknowledgments}
Y.~K., L.-P.~Z. and P.-M.~Z. are supported by the National Key Research and Development Program (Grant number is 2016YFE0130800) and the National Natural Science Foundation of China (Grant numbers are 11575254 and 11805242). This work is partly supported by Chinese Academy of Sciences President's International Fellowship Initiative (Grant numbers are 2018VMA0030 and 2018PM0028).



\begin{thebibliography}{99}  
\bibitem{Landau}
L.D.~Landau, 
{\it Z.~Phys.}~\textbf{64}, 629 (1930).
\bibitem{JJSakurai} J.~J.~Sakurai, J.~Napolitano,
{\it Modern Quantum Mechanics},
2nd edn.  (Addison-Wesley, San Francisco, 2011).
\bibitem{Lcan_Nature}
P.~Schattschneider et al.,
{\it Nature Comm.} \textbf{5}, 4586 (2014).  
\bibitem{Bliokh_PRX}
K.~Y.~Bliokh, P.~Schattschneider, J.~Verbeeck, and F.~Nori,
{\it Phys.~Rev.~X} {\textbf 2}, 041011 (2012).
\bibitem{Greenshields}
C.~Greenshields, R.~L.~Stamps, S.~Franke-Arnold, and S.~M.~Barnett,
{\it Phys.~Rev.~Lett.} \textbf{113}, 240404 (2014).
\bibitem{Nspin_review1}
E.~Leader, C.~Lorc\'{e},
{\it Phys.~Rep.} \textbf{541}, 163 (2014).
\bibitem{Nspin_review2}
M.~Wakamatsu,
{\it Int.~J.~Mod.~Phys.~A} \textbf{29}, 1430012 (2014).
\bibitem{path_dep_Wakamatsu}
M.~Wakamatsu, Y.~Kitadono, and P.-M.~Zhang,
{\it Ann.~Phys.} \textbf{392}, 287 (2018).
\bibitem{DeWitt}
B.~S.~DeWitt,
{\it Phys. Rev.} \textbf{125}, 2189 (1962).
\bibitem{KM1}
G.~Konstantinou, K.~Moulopoulos,
{\it Eur.~J.~Phys.}~\textbf{37}, 065401 (2016).
\bibitem{KM2}
G.~Konstantinou, K.~Moulopoulos,
{\it Int.~J.~Theor.~Phys.} \textbf{56}, 1484 (2017).
\bibitem{Landau_Lifshitz} L.~D.~Landau, E.~M.~Lifshitz,
{\it Quantum Mechanics: Non-Relativistic Theory}, Course of Theoretical Physics, Vol.~3 (Pergamon, New York, 1977). 
\bibitem{Murayama} H.~Murayama,
{\sl 221A Lecture Notes: Landau Level},
\url{http://hitoshi.berkeley.edu/221a/landau.pdf}.
\bibitem{JohnsonLippmann1}
M.~H.~Johnson and B.~A.~Lippmann,
{\it Phys.~Rev.} \textbf{76}, No.~6 828 (1949).
\bibitem{JohnsonLippmann2}
M.~H.~Johnson and B.~A.~Lippmann,
{\it Phys.~Rev.} \textbf{77}, No.~5 702 (1950).
\bibitem{van_Enk} S.~J. van Enk, 
to appear in {\it Am. J. Phys.}; arXiv:1906.00342.
\bibitem{CHLi_QWang}
C.-F. Li, Q. Wang,
{\it Physica B} \textbf{269}, 22 (1999).
\bibitem{VGW}
I.~D.~Vagner, V.~M.~Gvozdikov, and P.~Wyder,
{\sl Quantum mechanics of electrons in strong magnetic field},
HIT Journal of Science and Engineering, Vol.~3, Issue 1, 5 (Holon Institute of Technology, Holon, 2006).
\bibitem{Greenshields2} C.~Greenshields, S.~Franke-Arnold, and  R.~L.~Stamps, 
{\it New. J. Phys.} \textbf{17}, 093015 (2015). 
\bibitem{Schachinger} T.~Schachinger, S.~L\"{o}ffler, M.~St\"{o}ger-Pollach, P.~Schattschneider,
{\it Ultramicroscopy} \textbf{158}, 17 (2015).
\bibitem{Wakamatsu_PLA}
M.~Wakamatsu, Y.~Kitadono, L.-P.~Zou, and P.-M.~Zhang, {\it Phys. Lett. A}~\textbf{384}, 126415 (2020). 
\end{thebibliography}
\end{document}